\documentclass[12pt,preprint]{aastex}
\usepackage{gensymb} 
\usepackage[version=4]{mhchem}
\usepackage{amsmath}
\usepackage[colorlinks=true,linkcolor=blue,citecolor=blue]{hyperref}%
\shortauthors{Mandal \& Pal}

\begin{document}

\title{Detection of Low-Frequency QPO From X-ray Pulsar XTE J1858+034 During Outburst in 2019 with NuSTAR}

\author{Manoj Mandal}

\affil{Midnapore City College, Kuturia, Bhadutala, West Bengal, India 721129} 

\and

\author{Sabyasachi Pal}

\affil{Indian Centre for Space Physics, 43 Chalantika, Garia Station Road, Kolkata, India 700084\\
Midnapore City College, Kuturia, Bhadutala, West Bengal, India 721129} 

\begin{abstract}
\noindent
We study the timing properties of XTE J1858+034 using the Nuclear Spectroscopic Telescope Array (NuSTAR) and Burst Alert Telescope onboard Swift during the outburst in October--November 2019.	We have investigated for Quasi-Periodic Oscillation (QPO) during the outburst and detected a low-frequency QPO at $\sim$196 mHz with $\sim$6\% RMS variability from the NuSTAR observation. The QPO is fitted and explained with the model -- power law and a Lorentzian component. We have also studied the variation of QPO frequency with energy. The beat frequency model and Keplerian frequency model both are suitable to explain the origin of the QPOs for the source. Regular pulsations and QPOs are found to be stronger in high energy which suits the beat frequency model. The variation of the hardness ratio is studied over the outburst which does not show any significant variation.
\end{abstract}


\section{Introduction}
\label{intro}
In February 1998, the hard X-ray transient XTE J1858+034 was discovered by \citet{Ra98} using All-Sky Monitor (ASM) onboard Rossi X-ray Timing Explorer (RXTE) during an outburst. Series of observations were made after its discovery with the Proportional Counter Array (PCA) onboard RXTE, and frequent pulsation with a period of $221.0 \pm 0.5$ s was observed \citep{Ta98, Pa98}. The pulse profile with a pulse fraction of $\sim$25\% was found to be single-peaked and sinusoidal during the 1998 outburst \citep{Ta98, Pa98}. From the timing analysis, the presence of Quasi-Periodic Oscillation (QPO) at 110 mHz was discovered during 1998 outburst with around 7\% root mean square (RMS) fluctuation \citep{Pa98} and did not find any correlation between instantaneous X-ray flux with QPO frequency. \citet{Pa98} also looked for the most probable model to explain the origin of the QPO and suggested that both the beat frequency model and Keplerian frequency model are suitable to explain the origin of the QPOs for the source. They found the regular pulsations and QPOs are stronger in higher energy range which supports the beat frequency model. There was a second outburst in 2004 March and again in April which was first detected with Integral \citep{Mo04} and also followed up with RXTE. During April-May 2004 a strong outburst was detected using Integral \citep{Do08}, the hardness ratio did not show any significant variation during the outburst and the spin period was measured to be 220.4 $\pm$ 0.3 s with a sinusoidal shaped pulse profile. \citet{Mu06} observed a variable low-frequency QPO in the frequency range 140--185 mHz during the 2004 outburst from RXTE/PCA observations. They also studied the evolution of the QPO properties during the outburst using RXTE observations and observed significant energy dependence of the RMS variation in the QPOs. They did not found any correlation between QPO centroid frequency and the X-ray flux. They also found a positive correlation of pulse fraction with energy. The energy spectrum was well explained with a two components model -- a broken power law and blackbody emission along with a Gaussian at 6.4 keV. 

Recently the source went through an outburst in October 2019 detected by Gas Slit Camera (GSC) onboard MAXI \citep{Na19} and Burst Alert Telescope (BAT) on board Swift \citep{Ts21, Ma21}. We focus on the recent outburst and study the evolution of the timing properties during this time. During the recent outburst, the presence of low-frequency QPO is detected. The evolution of QPO frequency and its RMS variability is also studied. 

The data reduction and analysis methods are discussed in Section \ref{obs}. We have summarized the results of the current study in Section \ref{res}. The discussion and conclusion are summarized in the Section \ref{dis} and \ref{con} respectively.

\begin{figure}[t]
\centering{
\includegraphics[width=10cm]{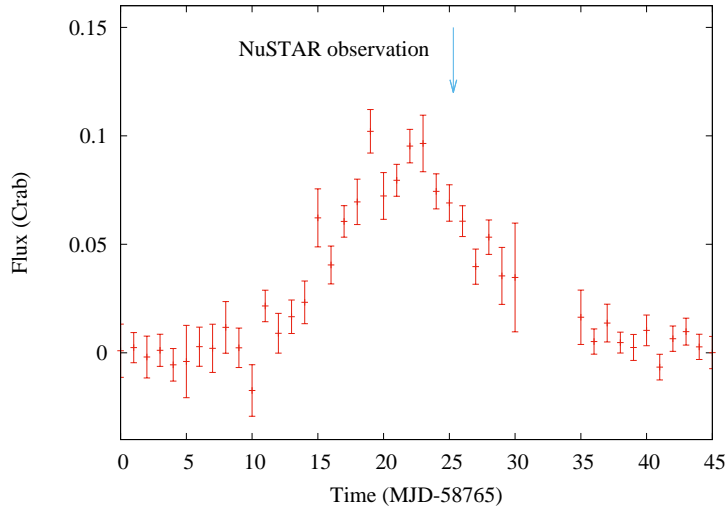}
	\caption{Flaring activity of XTE J1858+034 detected by Swift/BAT (15--50 keV) during the recent outburst (October--November 2019). The blue arrow shows the time of NuSTAR observation.}
\label{fig:BAT}}
\end{figure}

\section{Observation and data analysis}
\label{obs}
 We followed up the evolution of the outburst using Swift/BAT. We analyzed the NuSTAR data close to the peak of the 2019 outburst. We used the {\tt HEASOFT} v6.27.2 for the data reduction and analysis. 

 \begin{table}
\centering
\caption{Log of NuSTAR observation}
\label{tab:log_table}
\begin{tabular}{|c|c|c|c|c|} 
	\hline
	Instrument & Start time       & Date & Exposure  & Observation Id\\
		   &   (MJD)          & (yyyy-mm-dd)& (ks) &                    \\
\hline
NuSTAR & 58790.29754 & 2019-11-03 & 43.68 & 90501348002\\
\hline
\end{tabular}
\end{table}
 
NuSTAR was launched On the 13th of June 2012. The observatory comes with two co-aligned, identical X-ray telescope systems operating in a wide energy range from 3--79 keV. Separate solid-state CdZnTe pixel detector systems in each telescope usually referred to as Focal Plane Modules A and B (FPMA and FPMB; \citet{Ha13}), have a spectral energy resolution of 400 eV (FWHM) at 10 keV. NuSTAR performed an observation of XTE J1858+034 close to the peak of the outburst in the declining phase (MJD 58790) with a total exposure time of 43.68 ks (Obs. Id -- 90501348002). The observation log of NuSTAR is shown in the Table \ref{tab:log_table}. The data is reduced using the {\tt NuSTARDAS} pipeline version v0.4.7 (2019-11-14) provided under {\tt HEASOFT} v6.27.2 with latest {\tt CALDB}. We have extracted clean event files using {\tt DS9} version 8.1. We have generated the light curves of the source and background from circular regions centering the source with radii of 50\arcsec and 90\arcsec using {\tt NUPRODUCTS} scripts provided by the {\tt NuSTARDAS} pipeline. The {\tt lcmath} tool is used to combine light curves of the NuSTAR modules to improve the statistics for the timing analysis. 

BAT onboard the Swift observatory \citep{Ge04} is sensitive in hard X-ray (15-50 keV) consists of array of CdZnTe detectors \citep{Kr13}. We have used the results of the BAT transient monitor during the outburst, which is provided by the BAT team.


The light curves have been extracted from both the FPMA and FPMB science event data in different energy ranges with different bin sizes (0.01 s, 0.1 s, 1 s, and 10 s). Basic filtering criteria and corrections are applied to get clean continuous science event data. 
We have extracted the power density spectrum of the source to detect the presence of QPO. Power density spectrum analysis has been done using the {\tt POWSPEC} routine of {\tt FTOOLS} v6.27.2 which follows the FET algorithm. Power density spectrum with bin time 0.01 s is created using event mode data in the energy range 3--79 keV.

\section{Results}
\label{res}
The X-ray pulsar XTE J1858+034 went through an outburst detected by Swift/BAT\footnote{\url{https://swift.gsfc.nasa.gov/results/transients/}} with a maximum flux $\sim$0.1 Crab on MJD 58784. Figure \ref{fig:BAT} shows the variation of flux during the outburst using Swift/BAT (15--50 keV). The total duration of the outburst was around 3 weeks which started in the last week of October 2019 (MJD $\sim$ 58777) and continued till the second week of November 2019 (MJD $\sim$ 58800). We have summarized the result of the timing analysis of XTE J1858+034 during the recent outburst in October-November 2019. 

\subsection{Power density spectrum}
\begin{figure}[t]
\centering
\includegraphics[width=8.0cm,angle=270]{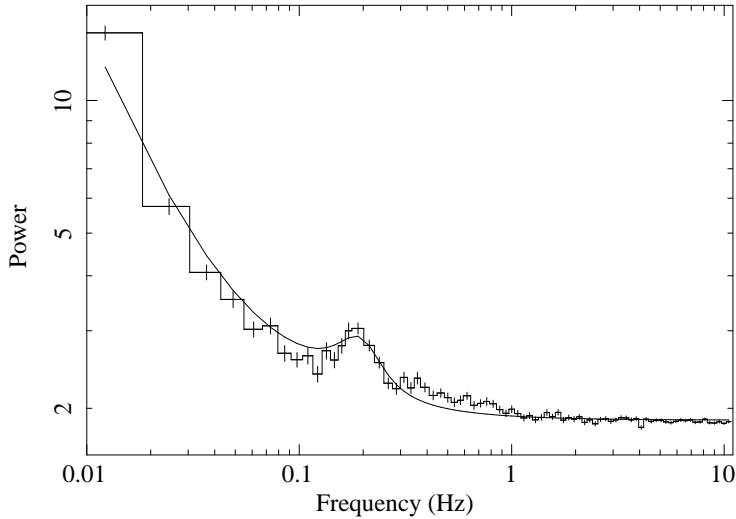}
\caption{The power spectrum of XTE J1858+034 generated from the 0.01 s binned light curve over the energy band 3--79 keV. The presence of QPO is observed at 0.196 Hz. The line represents the best-fitted model in frequency range 0.01--10.0 Hz consisting of a constant, power-law and a Lorentzian centered at the QPO frequency.}
\label{fig:powspec}
\end{figure}

QPOs have been detected for different accretion-powered X-ray pulsars in the range 1 mHz -- 40 Hz \citep{Ps06}. We have generated a Power Density Spectrum (PDS) with 0.01 s bin time light curves for both the module of NuSTAR -- FPMA and FPMB independently. A QPO feature is prominent at 0.196 Hz with RMS variability of 5.2\% for FPMA and for FPMB the QPO is prominent at 0.197 Hz with 6.3\% RMS variability. The power density spectrum with prominent QPO is shown in Figure \ref{fig:powspec} along with the best-fitted model using NuSTAR/FPMA data. The QPO feature is well fitted with a constant, a power-law component (with a power-law index -1.08), and a Lorentzian. QPO feature centered nearly 196 mHz for NuSTAR/FPMA with width (LW) 0.06 Hz and normalization (LN) 0.75.  Significant energy dependence of the RMS variation in the QPOs was observed in the previous outburst \citep{Mu06}. We have also investigated the energy dependence of the QPO feature. To study the energy dependence of QPO, we generated PDS in different energy ranges. The QPOs in the energy range 30--50 keV and beyond this range is not significant. We have not found any significant variation of the QPO frequency with energy.

\section{Discussion}
\label{dis}
We present the results of timing analysis using the NuSTAR data collected during the recent outburst of XTE J1858+034 in 2019. From the timing analysis, the spin period of the X-ray pulsar has been found $218.38 \pm 0.01$ s.  
We study the hardness ratio using the ratio of the count rates with time from Swift/BAT (15--50 keV) and MAXI/GSC (2--20 keV). We do not observe any significant change in the hardness ratio during the outburst. A low-frequency QPO has been detected at 0.196 Hz from the power spectrum during the 2019 outburst. It is well fitted by a constant, power-law, and a Lorentzian component. During the 2004 outburst, the QPOs were also detected in the frequency range 140--185 mHz. The QPO frequency in the recent outburst does not vary significantly with energy. Several models are used to explain the origin of QPOs \citep{Le88}. Generally, the beat frequency model \citep{Al85}, the Keplerian frequency model \citep{Va87}, and the magnetic disc precession model \citep{Sh02} are used to explain the QPOs in X-ray pulsars. For XTE J1858+034, the QPO frequency ($\nu_q$) is 196 mHz and spin frequency ($\nu_s$) is 4.58 mHz. As $\nu_q$ $>$ $\nu_s$ for this source, so the Keplerian frequency model may be applicable to explain the origin of QPO. According to this model, the QPOs are generated because of inhomogeneities in the Keplerian disk which attenuate the pulsar beam frequently. As the pulsations and QPOs are stronger in high energy which favors the beat frequency model, which is similar as observed during the 1998 outburst \citep{Pa98}. The beat frequency model implies that QPOs are generated due to the beat phenomena between the spin of the neutron star and the rotation of the innermost part of the disk \citep{Pa98}. Earlier both the Keplerian frequency model and beat frequency model were proposed to explain the origin of QPOs for this source \citep{Pa98} during the 1998 outburst and the pulsations and QPOs were found to be stronger in high energy, which favors the beat frequency model. Later in the 2004 outburst, the energy dependence of RMS variability of QPO and dependence of QPO frequency on X-ray flux was observed, which was more suitable to explain using the beat frequency model \citep{Mu06}.

\section{Conclusion}
\label{con}
We report a low-frequency QPO for the X-ray pulsar XTE J1858+034 at 0.196 Hz with $\sim$6\% RMS variability during the 2019 outburst. The QPO frequency does not show any significant variation with energy. The power spectrum is well described with the power-law and Lorentzian components. The origin of QPO for the source can be explained with the beat frequency model. The hardness ratio does not show any significant variation over the outburst.

\section*{Acknowledgements}
This research has done using data collected by NuSTAR, a project led by Caltech,  managed by NASA/JPL and funded by NASA, and has utilized the {\tt NUSTARDAS} software package, jointly developed by the ASDC (Italy) and Caltech (USA).


\begin{thebibliography}{56}
\bibitem[{Alpar \& Shaham} (1985)] {Al85}
Alper M. A., \& Shaham J., 1985, Nature, 316, 239

\bibitem[{Doroshenko} et~al.(2008)] {Do08}
Doroshenko F. R., Doroshenko A. V., Postnov A. K., Cherepashchuk M. A., Tsygankov S. S., 2008, Astron. Rep. 52, 138

\bibitem[{Gehrels} et~al.(2004)] {Ge04}
Gehrels N., et al., 2004, ApJ, 611, 1005

\bibitem[{Harrison} et~al.(2013)]{Ha13}
Harrison F. A. et al., 2013, ApJ, 770, 103

\bibitem[{Krimm} et~al.(2013)]{Kr13}
Krimm H. A. et al., 2013, ApJS, 209, 114

\bibitem[{Lewin} et~al.(1988)]{Le88}
Lewin W. H. G., van Paradijs J., van der klis M., 1988, Space Sci. Rev., 46, 273

\bibitem[{Malacaria} et~al.(2021)]{Ma21}
Malacaria C. et al., 2021, arXiv:2101.07020v1 [astro-ph.HE] 

\bibitem[{Mukherjee} et~al.(2006)]{Mu06}
Mukherjee U., Bapna S., Raichur H., Paul B., Jaffrey S. N. A., 2006, JApA, 27, 25

\bibitem[{Molkov} et~al.(2004)] {Mo04}
Molkov S. V. et al., 2004, ATel, 274, 1M

\bibitem[{Nakajima} et~al.(2019)] {Na19}
Nakajima M. et al., 2019, ATel, 13217, 1

\bibitem[{Paul \& Rao}(1998)]{Pa98}
Paul B, Rao A. R., 1998, A\&A, 337, 815

\bibitem[{Psaltis} (2006)]{Ps06}
Psaltis D., 2006, in Lewin W. H. G., van der Klis M., eds, Compact Stellar X-ray Sources. Cambridge Univ. Press, Cambridge, p. 1

\bibitem[{Remillard \& Levine} (1998)] {Ra98}
Remillard R., Levine A., 1998, IAUC, 6826, 1

\bibitem[{Tsygankov} et~al.(2021)]{Ts21}
Tsygankov S. S. et al., 2021, arXiv:2101.07030v1 [astro-ph.HE]  

\bibitem[{Shirakawa \& Lai} (2002)] {Sh02}
Shirakawa A., Lai D., 2002, ApJ, 565, 1134

\bibitem[{Takeshima} et~al.(1998)]{Ta98}
Takeshima T., Corbet R. H. D., Marshall F. E., Swank J., Chakrabarty D., 1998, IAUC, 6826, 1

\bibitem[{van der Klis} et~al. (1987)] {Va87}		
van der Klis M., Stella L., White N., Jansen F., Parmar A. N., 1987, ApJ, 316, 411

\end{thebibliography}
\end{document}